\documentclass[final]{elsart}
\usepackage{epsfig}
\usepackage{color}
\usepackage{amsmath}
\usepackage{amsfonts}
\usepackage{graphicx}
\usepackage{epstopdf}
\usepackage{ulem}

\usepackage[mathscr]{euscript}

\renewcommand{\thefootnote}{\arabic{footnote}}

\begin{document}

\begin{frontmatter}

\title{On the analysis of photo-electron spectra}
\author{C.-Z. Gao\corauthref{cor}$^{a,b}$},
\author{P. M. Dinh$^{a,b}$, P.-G.~Reinhard$^c$, E.~Suraud$^{a,b}$}
\corauth[cor]{Corresponding author\\{\it Email-address}~:
  gao@irsamc.ups-tlse.fr}

\address{$^a$Universit\'e de Toulouse, UPS; Laboratoire de Physique
             Th\'{e}orique (IRSAMC), F-31062 Toulouse Cedex, France}
\address{$^b$CNRS; LPT (IRSAMC), Universit\'e de Toulouse, F-31062 Toulouse Cedex, France}
\address{$^c$Institut f{\"{u}}r Theoretische Physik, Universit{\"{a}}t Erlangen,
            D-91058 Erlangen, Germany}

\date{2. May 2015}
\begin{abstract}
We analyze Photo-Electron Spectra (PES) for a variety of excitation
mechanisms from a simple mono-frequency laser pulse to involved
combination of pulses as used, e.g., in attosecond experiments.  In
the case of simple pulses, the peaks in PES reflect the occupied
single-particle levels in combination with the given laser frequency.
This usual, simple rule may badly fail in the case of
  excitation pulses with mixed frequencies and if resonant
  modes of the system are significantly excited. We thus develop
  an extension of the usual rule to cover all possible excitation
  scenarios, including mixed frequencies in the attosecond
  regime. We find that the spectral distributions of dipole,
monopole and quadrupole power for the given excitation taken together
and properly shifted by the single-particle energies provide a
pertinent picture of the PES in all situations. This leads to the
  derivation of a generalized relation allowing to understand
  photo-electron yields even in complex experimental setups.
\end{abstract}


\end{frontmatter}

\section{Introduction}
\label{sec:introduction}

Photo-Electron Spectroscopy (PES) has grown over decades to one of the
major tools of analysis of the structure and dynamics of atoms,
molecules or solids \cite{Tur62a,Tur70aB,Mai91b}.  With the
increasing availability and versatility of light sources, studies of
PES are now found in all areas of molecular physics, in atoms and
simple molecules \cite{Tje90a} as well as in more complex systems such
as clusters \cite{Hof01} or organic molecules \cite{Liu98a}.  The
archetypal case is that a system is subject to a laser field with moderate
intensity and sufficiently high frequency, such that all valence
electron states can be ionized in one stroke.  In this one-photon
regime, the PES delivers a printout of the sequence of single particle
(s.p.) energies from the occupied states \cite{McH89,Rab77aB}. But PES
do also allow to access dynamical features.  A typical example is the
multiphoton regime in which ionization proceeds via absorption of
several photons which is often achieved by IR photons at moderate
intensity. The PES then exhibits the pattern of Multiphoton
Ionization (MPI) \cite{Mai91b,Fai87,Fen10}
showing successive copies of the
s.p. spectrum. Further increasing laser intensity leads to
increasingly complex patterns, i.e., above threshold ionization (ATI) \cite{Ago79,Ebe91} and strong-field ionization \cite{DeW98,Cam01,Hat07}.

Simple laser setups are characterized
by {\it one} well defined photon frequency $\omega_\mathrm{las}$. The
energies $E_{i,\nu}$ of the PES peaks from such sources follow a well known
rule relating them to s.p. energies $\varepsilon_i$ \cite{Fai87} as
\begin{equation}
E_{i,\nu} = \varepsilon_i + \nu \hbar \omega_\mathrm{las} \quad, 
\label{eq:basic}
\end{equation}
where $i$ stands for the s.p. state from which the electron is
emitted and $\nu$ is the number of absorbed photons for ionization
($\nu>1$ for MPI). 
In other words, the PES yield $\mathcal Y \left(E_\mathrm{kin}\right)$ is then interpreted as
a sum of Dirac distributions~:
\begin{equation}
\mathcal Y \left(E_\mathrm{kin}\right) \ \leftrightarrow \ \sum_\nu \sum_i \delta(E_\mathrm{kin} - E_{i,\nu}) \quad.
\label{eq:y_dirac}
\end{equation} 
For high laser intensity, relation (\ref{eq:basic})
has to be complemented by the energy of the ponderomotive potential
$U_p$ which represents the average kinetic energy of a free electron in
the laser field \cite{Fen10,Kel65}.
For the sake of simplicity, we shall confine our discussion in the
following to cases in which $U_p$ remains negligible.

With the advent of a large variety of coherent light sources
\cite{Kel03,Rul05,Pas08}, it has become possible to access more complex
dynamical scenarios, involving several frequencies and/or pulse
combinations, for example, in pump-and-probe setups \cite{Zew94}. The
latter ones even provide a time-resolved access to dynamics, at ionic pace
with femtosecond (fs) \cite{Sei00,And02} down to electronic pace with
attosecond (as) pulses \cite{Pau01,Kra09,Cor07,Joh05,Nei13}.  
The PES of course remains a highly valuable tool of investigation of the
dynamics, as it basically provides insight into electronic dynamics
via ionization characteristics.  
But the simple rule Eq.~(\ref{eq:basic}) takes a more complicated
expression in the case of a complex light pulse. For instance, in a
pump-and-probe experiment where there are possibly two laser colors,
one should add to the s.p. energies linear combinations of the two
laser frequencies to obtain the positions of the PES peaks.  This is
all the more involved when it is in the multi-photon regime
for which the number of combinations grows fast. In this case, the
identification of the PES peaks to a certain combination of laser
frequencies can become tricky.  It is thus important to understand
how the PES builds up in the course of irradiation processes involving
complicated laser pulses.  The question is in fact even more general.
There also exist experiments measuring kinetic energies of the
electrons emitted after irradiation by a charged projectile
\cite{Kel10,Nan13} and a first theoretical exploration was proposed in
\cite{Din15}. This also provides a PES but now with a ``photon'' of
mixed frequencies, as the effect of a charged projectile is basically
to deliver a short electromagnetic pulse which covers a broad band of
frequencies. Indeed, the shorter the pulse, the broader the band
  of accessible frequencies. For the sake of simplicity, we continue
to call PES such kinetic electron spectra from ultra-short pulses,
even if the notion of a ``photon'' is rather untypical here.  The
analysis of the PES obtained in such fast collisions can also reveal
structures which can be attributed to eigenfrequencies of the system,
e.g., the Mie plasmon frequency in the case of metal clusters
\cite{Hee93,Bra93} or C$_{60}$ \cite{Her06aR}.  It should also be
noted that the intense plasmon of metal clusters also shows up in
laser-driven PES and can be spotted when scanning laser frequency
around the plasmon frequency, as was outlined some years ago
\cite{Poh01}.

All in all, the PES thus reflects the typical frequencies present in
the system, either those delivered from outside by a laser field or
intrinsic ones as mostly visible in the case of a fast collision.  It is
the goal of this paper to investigate in detail this interplay of
frequencies.  We shall consider various types of excitations in order
to scan a variety of dynamical scenarios. At the side of test systems,
we perform the quantitative analysis mostly for a simple case, that is the He atom,
to maintain clear signatures. 
The paper is organized as follows.  Section~\ref{sec:theory} briefly presents
the theoretical framework. Then we analyze a few typical cases involving various laser setups
including state-of-the-art attotrains in Section
\ref{sec:testcases}. Once identified the limitations of the simple
rule Eq. (\ref{eq:basic}), we propose an alternative, more
general, rule to cover any dynamical scenario in Section
\ref{sec:quad}. Conclusions and perspectives are finally drawn in
Section \ref{sec:conclusion}.

\section{A brief sketch of theory}
\label{sec:theory}

\subsection{Basics}

Our computations are done with Density Functional Theory
(DFT).  We work here in real-time Time-Dependent DFT at the level of
the Kohn-Sham (KS) picture \cite{Koh65}. The system is then described by a
set of single particle wave functions $\varphi_i({\bf r},t)$ which
follow the time-dependent Kohn-Sham (TDKS) equations \cite{Gro90}
\begin{equation}
\mathrm{i}\frac{\partial \varphi_i({\bf r},t)}{\partial t} = 
\left[-\nabla^2 + V_\mathrm{eff}({\bf r})\right]\varphi_i({\bf r},t).
\label{eq:tdks}
\end{equation}
We use Rydberg atomic units ($\hbar=e=2m_e=1$) throughout the
paper except where specifically stated differently.  The term
$V_\mathrm{eff}$ in Eq. (\ref{eq:tdks}) is the KS effective
potential. It is composed of three parts $V_\mathrm{eff} =
V_\mathrm{ion} + V_\mathrm{H} + V_\mathrm{xc}[\varrho]$. The term
$V_\mathrm{ion}$ is the ionic background potential which is described
by a pseudopotential, a simple local one for sodium \cite{Kue98} and a
Goedecker-type one for helium \cite{Goe96}. The term $V_\mathrm{H}$ is
the Hartree contribution and the last term $V_\mathrm{xc}$ represents
the exchange correlation potential. The latter term is a functional of
the local density $\varrho({\bf r},t) = \sum_i |\varphi_i({\bf
  r},t)|^2$ and we work here within the time-dependent version of
  the Local Density Approximation (TDLDA) with the parametrization of
Perdew and Wang \cite{Per92}. The TDLDA is complemented by an
efficient self-interaction correction term \cite{Leg02} which allows
us to properly describe the ionization threshold~\cite{Klu13a,Wop15}.

\subsection{Excitation mechanisms}

In this paper, we will use various irradiation processes for the excitation~: a fast
colliding charged projectile, a single laser pulse, a sum of attosecond pulses, and finally 
an involved superposition of IR femtosecond pulse with UV attosecond pulses. 

The first type of excitation is an instantaneous dipole boost of
all electronic wave functions at $t=0$, mimicking the collision with a
swift ion, that is, $\varphi_i(\mathbf r,t=0)= e^{i\eta \hat{D}} \,
\varphi_i^{(0)} (\mathbf r)$, where $\hat{D}$ is the dipole operator,
$\eta$ is the boost momentum and $\varphi_i^{(0)}$ is the ground state
Kohn-Sham wave function for state $i$.

The single laser pulse is modeled as
\begin{subequations}
\label{eq:Elas}
\begin{eqnarray}
  E_\mathrm{las}(t)
  &=&
  E_0 \, f_{T}(t) \, \sin(\omega_\mathrm{las}t)
  \;,
\label{eq:simplepulse}
\\
  f_T(t)
  &=&
  \cos^2\left(\frac{t-T/2}{T}\pi\right)\theta(t)\theta(T-t) \quad,
\end{eqnarray}
\end{subequations}
where $\theta$ is the Heaviside function, $T$ the pulse duration and $\omega_{\rm las}$ its frequency. 
$T$ takes typical values of some hundreds of fs, while we will consider in this paper either an IR frequency 
($\omega_{\rm las}=\omega_{\rm IR}=0.115$~Ry) or a UV one ($\omega_{\rm las}=\omega_{\rm UV}=1.5$~Ry
in Figure~\ref{fig:XUVna9p} and 0.5~Ry in Figure~\ref{fig:mpi}). 

\begin{figure}[htbp]
\centerline{\includegraphics[width=0.85\linewidth]{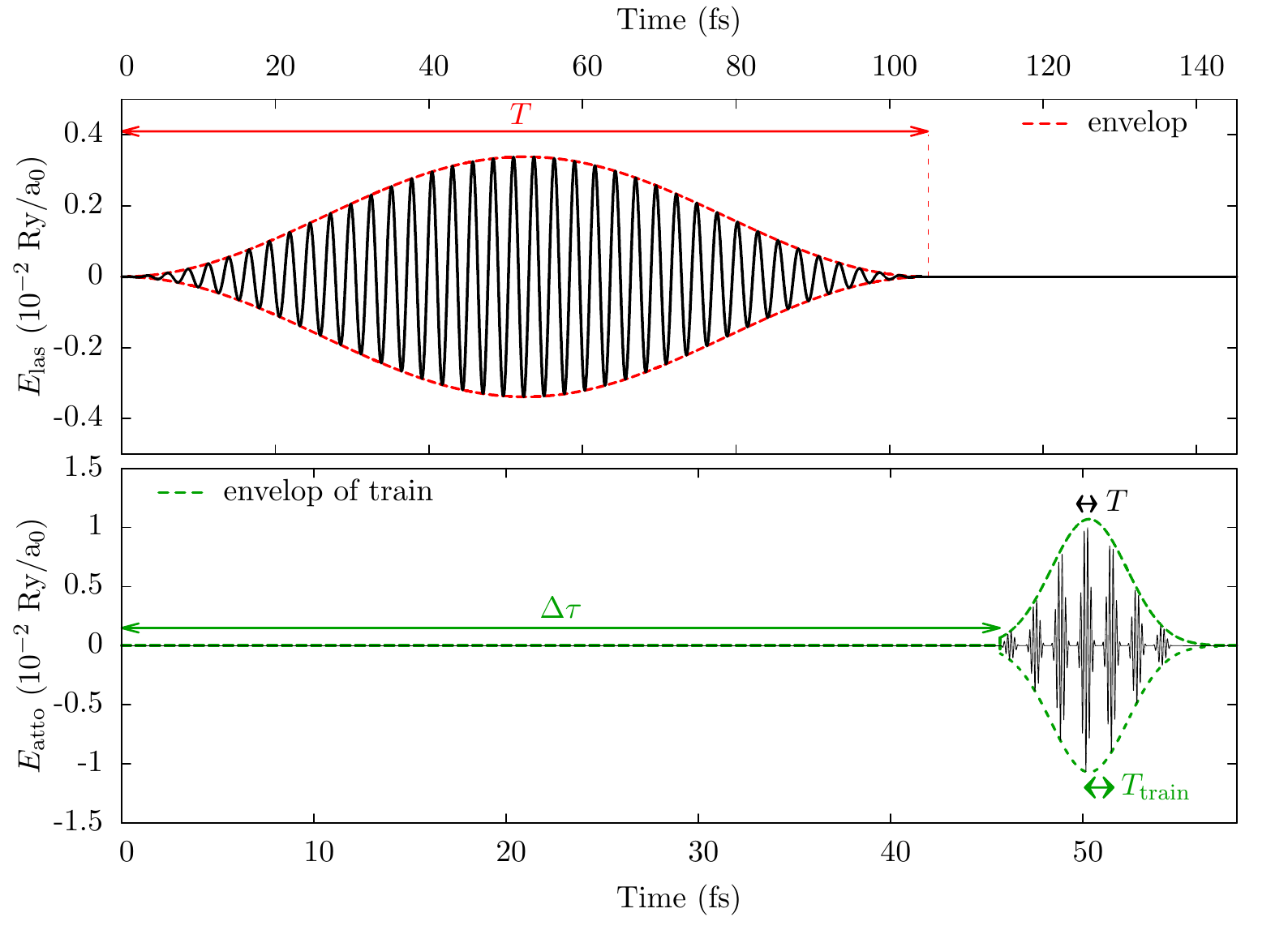}}
\caption{Time profiles of typical laser fields considered in this paper. Top~: IR femtosecond pulse
as defined in Eqs.~(\ref{eq:Elas}),
with duration $T=105$~fs, frequency $\omega_{\rm las}=0.115$~Ry, intensity $I=10^{13}$~W/cm$^2$. 
Bottom~: train of UV attosecond pulses (attotrain), as defined in Eqs.~(\ref{eq:attotrain}),
with $\omega_{\rm atto}=1.69$~Ry, $I=10^{12}$~W/cm$^2$, attopulse duration $T=1$~fs, 
delay $\Delta \tau=46$~fs, and attopulse separation $T_{\rm train}=1.33$~fs.}
\label{fig:laser}
\end{figure}

As an example of a rather complex pulse, we will consider an attosecond train of UV pulses, that is, a sum of pulses similar to that defined in
Eqs.~(\ref{eq:Elas}) but with $\omega_{\rm las}=\omega_{\rm atto}$ in
the UV range, a pulse duration $T$ of a fraction of fs, and each
UV attopulse shifted in time. In other words, the attotrain field
reads~:
\begin{subequations}
\label{eq:attotrain}
\begin{eqnarray}
  E_\mathrm{atto}(t)
  &=&
  E^\mathrm{(atto)}_0
  \sum_{\alpha=0}^{N-1}
  g(t)
  f_T(t-t_\alpha)\sin(\omega_\mathrm{atto} t)
  \;, \label{eq:Eatto} \\
  t_\alpha
  &=&
  \Delta \tau
  + \alpha \, T_{\rm train}
  \;, \label{eq:talpha}\\
g(t) &=&  \exp \left(- \frac{(t-\Delta \tau-\mathscr T)^2}{{\mathscr T}^2/(4 \ln 2)} \right)
\;, \label{eq:atto_envel}\\
\mathscr T &=&N \, T_{\rm train} /2 \; .
\end{eqnarray}
\end{subequations}
Some additional parameters have been introduced here, let us explain them.
In most experiments \cite{Joh05,Nei13,Joh07,Hol11,Klu11},
an IR fs pulse is used to generate high harmonics which then serve to produce 
the coherent attotrain, which is usually modulated by an envelop that we denoted by $g$ 
in Eq.~(\ref{eq:atto_envel}). Here, we take in our simulations a Gaussian envelop as in the 
experiment of~\cite{Nei13}. (But other theoretical calculations use a $\cos^2$ envelop instead
for a comparison with another experiment~\cite{Joh07}.) 
One can also play on the delay between the IR pulse and the attotrain.
This delay $\Delta \tau$ is thus entering the definition of the envelop $g$ but also in the shifting times
$t_\alpha$ in Eq.~(\ref{eq:talpha}). In the present work, we fix it at $\Delta \tau=46$~fs.
As for $T_{\rm train}$, it corresponds to the time separation
between the maxima of two successive attopulses.
Finally, the number of UV pulses in the attotrain is denoted by $N$
and is typically about 10.
It is obvious that Eq.~(\ref{eq:Eatto}) represents a considerably complicated pulse and
that it will provide a critical test case for our analysis. Figure~\ref{fig:laser} summarizes
the various pulses we will use, as it displays the time
profile of a single IR fs pulse (top panel), see Eqs.~(\ref{eq:Elas}), and that of
an attotrain (bottom panel), see Eqs.~(\ref{eq:attotrain}).

Finally, we will consider the superposition of an IR fs pulse and a UV attotrain, as 
encountered in experiments~\cite{Joh05,Nei13,Joh07,Hol11,Klu11}. 
This case will represent the most difficult excitation mechanism to be understood.

\subsection{Numerical treatment}

The TDLDA equations are implemented on a grid in coordinate space.  In
the present (principle) study, we consider spherical systems so that
we can recur to a 2D cylindrical representation. This
considerably reduces the computational expense.  The ground state of
the system is determined by the damped gradient method. The time
propagation uses $T$-$V$ splitting. To describe ionization, we
use absorbing boundary conditions. For details of the numerics, see
\cite{Cal00,Rei03a}.  A warning is in order here. Experimental
  PES and related effects belong to electrons in the continuum. It is
known that a numerical description of outgoing waves requires careful
choice of grid parameters to avoid artifacts from unwanted
discretization of the continuum \cite{Rei06c}. We therefore choose
here very large numerical boxes to avoid such artifacts~: for the
  metal cluster ${{\rm Na}_9}^+$, we use 138.4~a$_0$ in the
  longitudinal direction ($=$symmetry axis) and 71.2~a$_0$ in the
  radial one (with the same mesh size of 0.8~a$_0$ in both
  directions), and for the He atom, we use 105.6~a$_0$ and
  $52.8$~a$_0$ respectively (the mesh size being in that case
  0.6~a$_0$).

\setcounter{footnote}{0}

\subsection{Observables}

The spectral distribution of multipole strength is computed from TDLDA
in the time domain using spectral analysis following the prescription
of \cite{Cal95a,Cal97b}.  For the example of dipole strength, this
proceeds as follows. Whatever the excitation mechanism (boost or
finite-width laser pulse), we record the dipole moment
$D(t)=\int\mathrm{d}^3 \mathbf r\,z\,\varrho(\mathbf{r},t)$ as it evolves over
the TDLDA simulation. Note that we consider here the dipole
along laser polarization axis, denoted by $z$, or, generally speaking, along the excitation
direction and/or symmetry axis of the system.  The time signal $D(t)$
is Fourier transformed into frequency domain, yielding
$\tilde{D}(\omega)$. The dipole power spectrum is then
$P_D(\omega)\propto\left|\tilde{D}(\omega)\right|^2$. We consider here
the power spectrum rather than the usual dipole strength (which is the
imaginary part of $\tilde{D}(\omega)$ \cite{Cal97b}) since we
compare it to the PES, which is also a power spectrum.

The central ingredient of the paper is the PES, the spectrum of
asymptotic kinetic energies of emitted electrons. Its computation has
been worked out in detail in several papers
~\cite{Poh01,Poh04a,Bae10,DeG12}~; for a recent extension to the case of
strong fields, see~\cite{Din13}. We briefly present the procedure for the sake
of completeness. It is directly based on the
KS s.~p. wave functions \cite{Poh01}. We choose
a ``measuring point'' $\mathbf{r}_\mathcal{M}$ located far away from
the system and just before the absorbing boundaries.  We record all
s.~p. wave functions $\varphi_i(\mathbf{r}_\mathcal{M},t)$
at that point all along the simulation time.  Because of the large
distance from the center of the system, we can neglect the KS field and
assume a free particle dynamics (strong fields, not encountered
here, require to consider electron motion in the ponderomotive field
of the laser \cite{Din13}).  In addition, because we are close to the
absorbing boundary and far away from the source, only outgoing waves
with momentum $\mathbf{k}=k\mathbf{r}_\mathcal{M}/r_\mathcal{M}$ will
pass the point $\mathbf{r}_\mathcal{M}$.  This allows us to establish a
revertible relation $\omega\leftrightarrow k$ between momentum and
energy. The PES yield
$\mathcal{Y}_{\Omega_{\mathbf{r}_\mathcal{M}}}(E_{\rm kin}) $ can then
be obtained from Fourier transformation from time $t$ to frequency
$\omega$ of the KS orbitals $\varphi_j$:
\begin{equation}
  \mathcal{Y}_{\Omega_{\mathbf{r}_\mathcal{M}}}(E_{\rm kin}) 
  \, \propto \, 
  \sum_{j=1}^N
  \left| \widetilde{\varphi}_j(\mathbf{r}_{\mathcal{M}},E_{\rm kin})\right|^2.
\label{eq:raw3D}
\end{equation}
The quantity $\Omega_{\mathbf{r}_\mathcal{M}}$ represents here the
solid angle related to the direction of $\mathbf{r}_\mathcal{M}$,
$\widetilde\varphi_j(\mathbf{r}_\mathcal{M},E_\mathrm{kin})$ is the
time-frequency Fourier transform of
$\varphi_j(\mathbf{r}_\mathcal{M},t)$ and the kinetic energy reads
$E_{\rm kin}=k^2/2 =\omega$ ~\cite{Din13}.  Performing such an analysis
at a dense mesh of measuring points allows us to compute the fully
energy- and angular-resolved PES \cite{Wop15,Wop12c}.  However, since we are
interested here purely in spectral features, we shall restrict
the analysis to a point along the laser polarization (or symmetry) axis.

\section{Typical PES interpretation and limitations thereof}
\label{sec:testcases}

\subsection{A simple example}
\label{sec:ex}

\begin{figure}[htb!]
 \centerline{\includegraphics[width=0.85\linewidth]{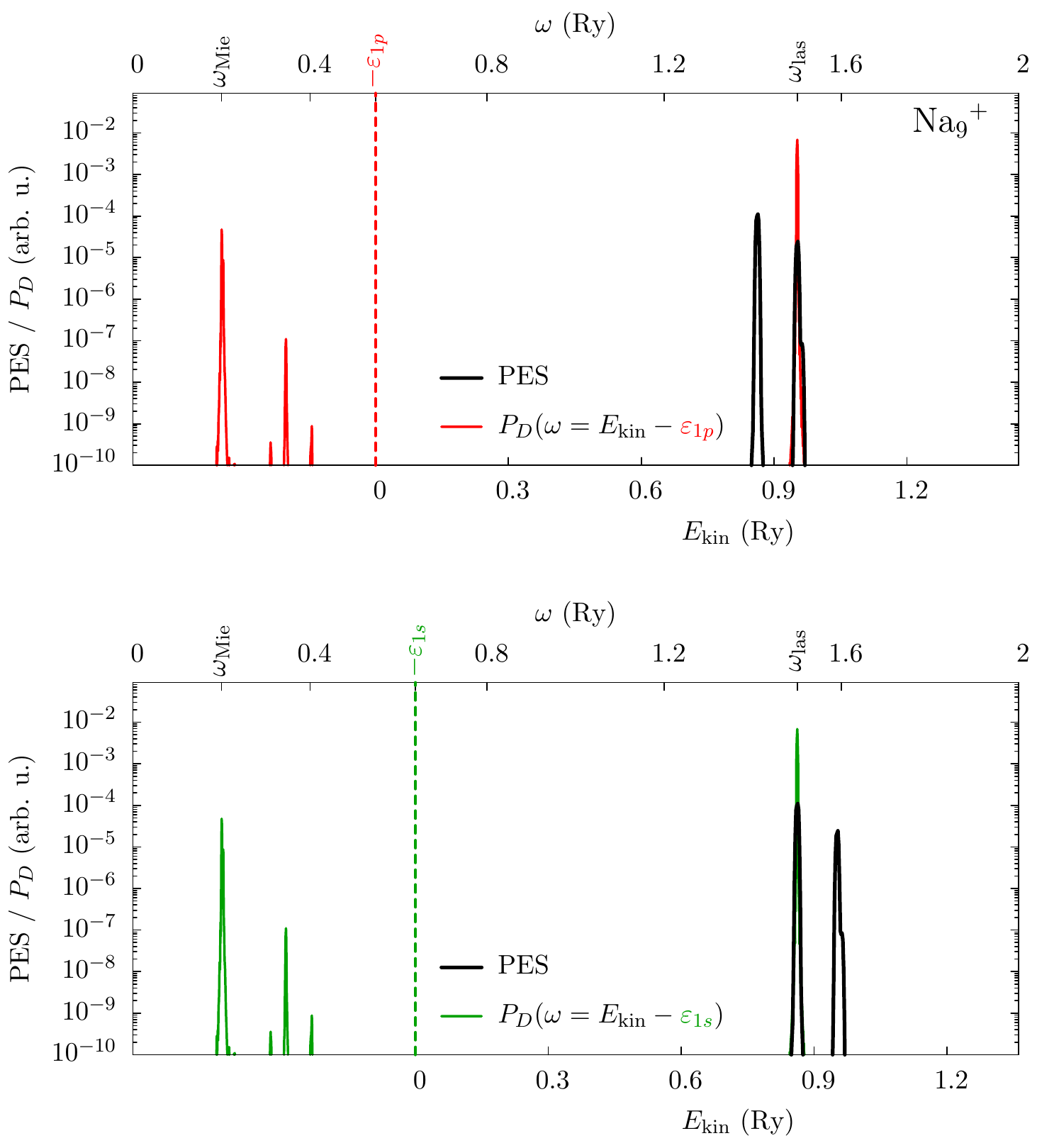}}
 \caption{Excitation of ${{\rm Na}_{9}}^+$ by a UV laser pulse with
   frequency $\omega_{\rm las}=1.5$ Ry, intensity of $10^{13}$
   W/cm$^{2}$, and total duration 232 fs. PES (black curve, lower $x$-axis in each panel) 
is compared with
   the spectral distribution of dipole power $P_D(\omega)$, one
   shifted by $\varepsilon_{1s}$ (bottom panel) and the other shifted by
   $\varepsilon_{1p}$ (top panel) according to Eq.~(\ref{eq:basic}). PES is drawn versus the kinetic energy
   $E_\mathrm{kin}$ of the emitted electron and exists, of course, only
   for positive $E_\mathrm{kin}$, while the dipole power $P_D$ shifted by
   $\varepsilon_i$ also extends to negative energies. The vertical dashed line indicates
   the emission threshold at $E_\mathrm{kin}=0$. PES and $P_D$ are drawn
   in arbitrary units and are scaled such that they have comparable
   peak heights. }
 \label{fig:XUVna9p}
\end{figure}

We start with illustrating Eq. (\ref{eq:basic}) through  a simple test
case, namely the ${{\rm Na}_{9}}^+$ cluster irradiated by a single UV laser pulse,
see Eq.~(\ref{eq:Elas}), of frequency $\omega_\mathrm{las} = 1.5$ Ry, intensity $I =
10^{13}~\rm {W/cm}^2$ and duration $T=232$ fs.  
The ponderomotive
potential is $U_p$= $2.2\times 10^{-4}$ Ry, thus negligible.  The
resulting total ionization is $N_{\rm esc}=0.038$, which means
that the test case stays safely in the perturbative one-photon regime.
The s.p. spectrum of the ${{\rm Na}_9}^+$ cluster groups into two shells of
nearly degenerated levels denoted in harmonic oscillator labeling as
$1s$ and $1p$ \cite{Bra93}.  Figure \ref{fig:XUVna9p} displays the
obtained PES (black line) for this laser pulse. 
The PES shows the typical pattern of one-photon emission.  One can
nicely identify the two peaks corresponding to the two shells, $1s$ and
$1p$, of occupied s.p. levels (mind that, in 2D, the $1p$ levels are
degenerate), shifted by the laser frequency, as expected from
Eq. (\ref{eq:basic}).

Figure \ref{fig:XUVna9p} also shows the
dipole power distribution $P_D(\omega)$ computed from the response
$D(t)$ to the given laser pulse. The power spectrum is compared with the PES
twice, once shifted by the s.p. energy $\varepsilon_{1s}=-0.64$~Ry (bottom panel) and 
once shifted by $\varepsilon_{1p}=-0.55$~Ry (top panel).  These
$P_D(\omega\!=\!E_\mathrm{kin}-\varepsilon_i)$ represent the
generalization of Eq.~(\ref{eq:y_dirac}) from one dominant laser
frequency $\omega_\mathrm{las}$ to a full spectrum $\omega$.
Let us now analyze in more detail $P_D(\omega)$ (we remind that the scale for $\omega$
appears as the upper horizontal scale in each panel). It shows two prominent peaks~: the leading, expected, one at
$\omega_{\rm las}$ arising from the laser field and a small
secondary peak at lower frequency $\omega_{\rm Mie}=0.2$ Ry
corresponding to the much celebrated Mie plasmon resonance
$\omega_{\rm Mie}$. As it is visible in the right parts of
each panel of Figure~\ref{fig:XUVna9p}, the laser peak of
$P_D$ shifted by the two $\varepsilon_i$ perfectly coincides with
the two PES peaks. One may be surprised by the appearance of a
  Mie resonance peak because the mismatch between
  $\omega_\mathrm{Mie}=0.2$~Ry and $\omega_{\rm las}=1.5$~Ry is
  huge. Remind however that the laser pulse has a finite width. Therefore, even at 
$\omega_\mathrm{Mie}$, there remains a faint piece of spectral strength
in the laser signal which together with the overwhelming responsivity
of the Mie plasmon produces this sidepeak. Although visible, it is
naturally suppressed by orders of magnitude and has no impact on the
PES here.

Thus far, the picture is straightforward for such a simple laser
pulse.  The analysis will nevertheless quickly grow in complexity when
proceeding to more involved pulses with a richer spectral pattern,
because it has to be performed with the two occupied electron shells
in the case of ${{\rm Na}_9}^+$.  To simplify the picture at the side
of the test system, we will concentrate, from now on, on the simpler
case of an helium atom with one single occupied level (with two
electrons of opposite spins). It should also be noted that the
  Helium case is especially interesting in our framework as it
  involves only one (doubly occupied) wave function. For then the self
  interaction correction which we use \cite{Leg02} exactly removes the
    self-interaction error at all orders. 
 
\subsection{Dipole boost excitation}

\begin{figure}[htbp]
\centerline{\includegraphics[width=0.85\linewidth]{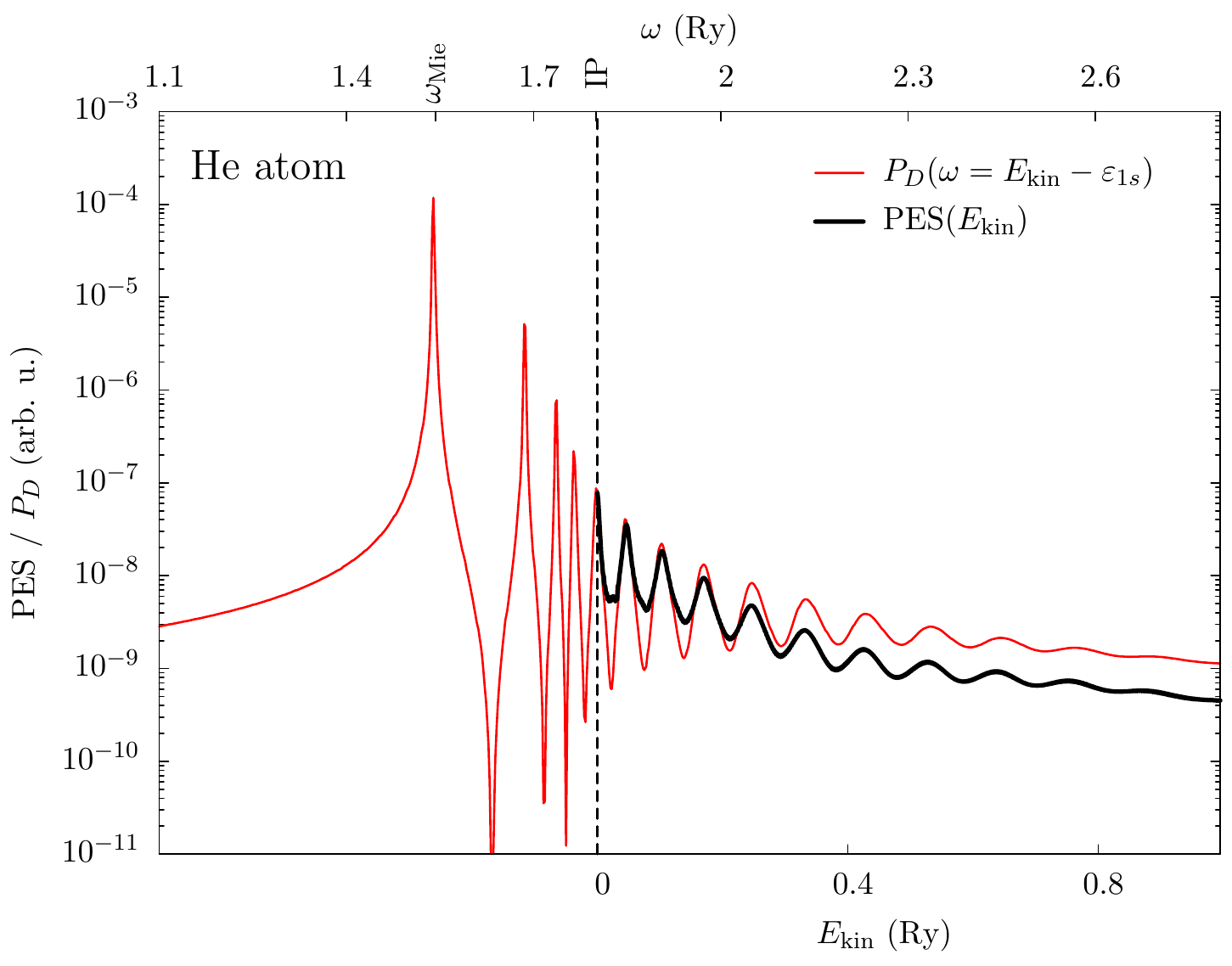}}
 \caption{He atom excited by
   instantaneous dipole boost of 0.001/a$_0$.
Photo-Electron Spectra (black line and lower horizontal scale on $E_{\rm kin}$) 
and dipole power $P_D$ shifted by
the energy of the s.p. level, $\varepsilon_{1 \rm s}=1.8$~Ry, (red line and upper horizontal scale
on $\omega$). The key for $P_D$ indicates that it can be read in two ways, either vs. $\omega$
   (upper $x$-axis) as usual for a spectrum, or vs.
   $E_\mathrm{kin}$ (lower $x$-axis) for comparison with PES.  The
   vertical solid line denotes the ionization potential (IP).}
 \label{fig:boostHe}
\end{figure}

We start the studies on the He atom by first considering a simple
boost excitation (simulating a collision with a fast charged
projectile).  Figure \ref{fig:boostHe} shows the emerging PES (black)
together with the dipole power spectrum $P_D$ (red).
In that case, the simple rule Eq.(\ref{eq:basic}) is by construction
not applicable because there is no laser around imprinting its
frequency $\omega_\mathrm{las}$ onto the process. On the other hand,
and following the analysis of Section~\ref{sec:ex}, one can still
compare the distribution of PES and the dipole spectrum
$P_D(\omega\!=\!E_\mathrm{kin}\!-\!\varepsilon_{1s})$, as is done in
Figure \ref{fig:boostHe}.  The sequence of peaks in PES nicely matches
a similar sequence in $P_D$ when shifting the scales by the
s.p. energy of the single occupied state $1s$ (which is here identical
to the IP). The agreement between both curves is striking, primarily
concerning the location of peaks and, to a large extent, even at the
side of the relative heights of peaks. This confirms the findings of
Figure \ref{fig:XUVna9p}, but now in a case where the simple rule
Eq. (\ref{eq:basic}) is by definition meaningless.
The generalization of the rule (\ref{eq:basic}) which we read off from
Figure \ref{fig:boostHe} is that the whole distribution of PES yield is
strongly related to the, properly shifted, dipole power, i.e., $\mathcal{Y}(E_\mathrm{kin})\leftrightarrow
P_D(\omega\!=\!E_\mathrm{kin}\!-\!\varepsilon_{1s})$, both to be read
as a function of $E_\mathrm{kin}$. This is the conjecture which we now want to
scrutinize further.

\subsection{An attotrain excitation}

\begin{figure}[htb!]
 \centerline{\includegraphics[width=0.85\linewidth]{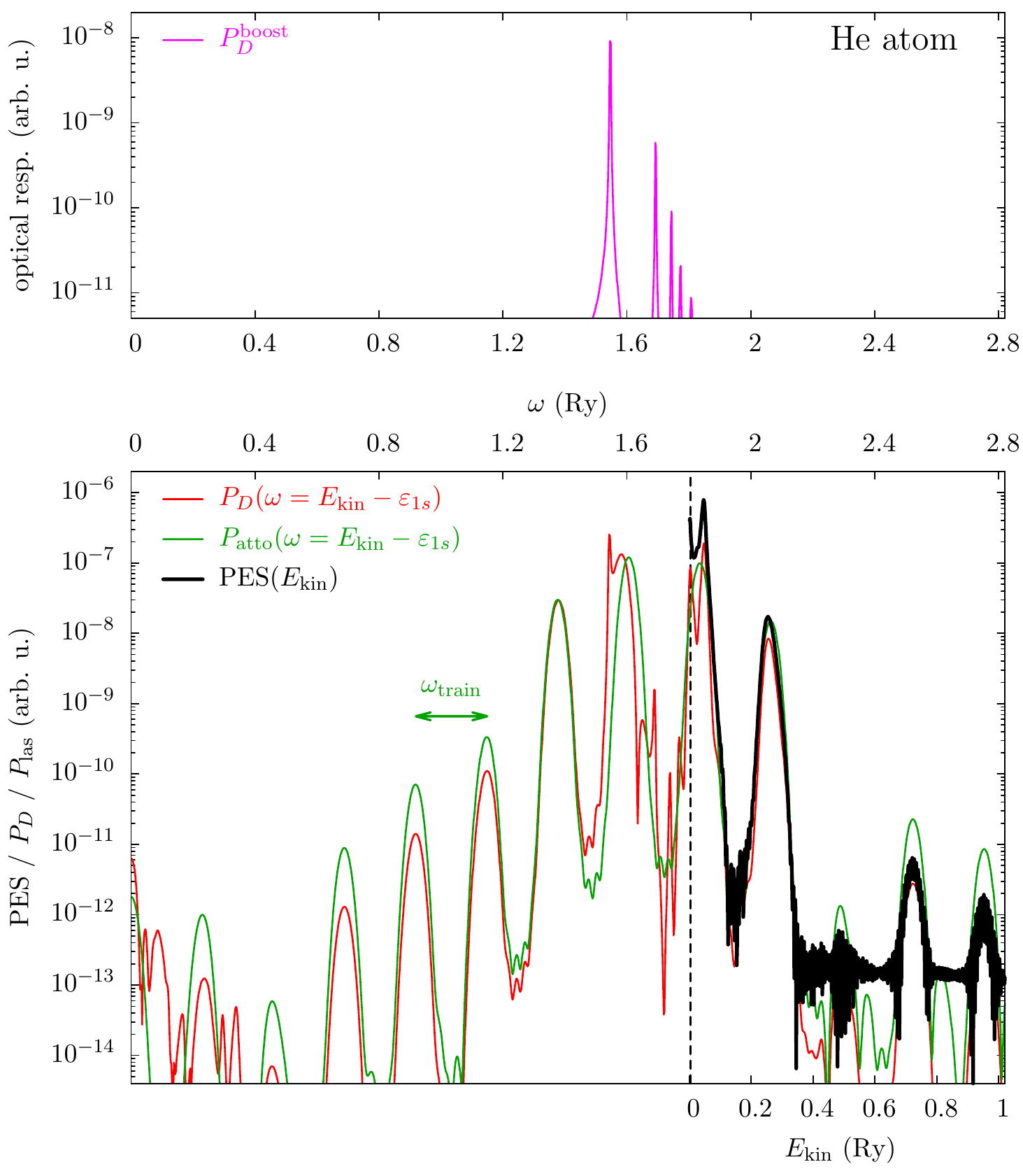}}
 \caption{Lower: PES (black) and dipole power $P_D$ (red) for a He
   atom excited by the attosecond pulse train (APT) as given in Eq.
   (\ref{eq:attotrain}), see text for details. The green line shows 
   the power spectrum $P_\mathrm{atto}$ of the laser field delivered by the
attopulse train.  The
   vertical dashed line indicated the ionization threshold.  Upper: The
   optical response $P_D^{\rm boost}$ of the He atom.}
 \label{fig:pureAPT}
\end{figure}

The case of a simple boost produces a relatively simple dipole
spectrum $P_D$ with only a few dominant peaks and minor ones (which,
nevertheless, show up in the PES). It is thus also interesting to see
how the above conjecture performs in a complicated case involving
several laser frequencies on top of the eigenfrequencies of the
system.
This situation is illustrated in Figure \ref{fig:pureAPT}, again in
the helium atom. The latter one is irradiated by the attosecond laser train defined in
Eqs.~(\ref{eq:attotrain}). 
It consists out of $N=7$ individual UV pulses, each one of
duration $T=1$ fs. The time interval between the end of an attopulse and the onset of the next one is 0.33 fs. Therefore, we have $T_{\rm train}=1.33$~fs, corresponding to a frequency $\omega_\mathrm{train}=2\pi/(1.33/0.0484)=0.23$~Ry to the
system.  The lower panel of Figure \ref{fig:pureAPT} shows the result
for PES and $P_D(\omega)$.  For the sake of completeness, we  also show 
the power spectrum of the electric field
$P_\mathrm{atto}(\omega)$ defined as $P_\mathrm{atto}(\omega)= |{\tilde E}_\mathrm{atto}(\omega)|^2$
where ${\tilde E}_\mathrm{atto}(\omega)$ is the Fourier transform of
the laser field of the attopulse train $E_\mathrm{atto}(t)$.
The dipole power spectrum $P_D$ now takes a seemingly complicated 
``comb\rq\rq{} structure which, however, can be understood in simple terms.  One
identifies a wide peak located around the attosecond frequency
$\omega_{\rm atto}=1.69$ Ry (see upper horizontal scale; mind that there is
a Gaussian envelop to the attotrain, see Eq.~(\ref{eq:Eatto}), so that 
the maximum of $P_{\rm atto}$ is not exactly at $\omega_{\rm atto}$). It has a large width of about 0.3 Ry
related to the short duration of only 1 fs. Because the  signal is
repeated in time (attotrain), this broad structure is overlayed by
more intense and sharper structures precisely separated from each
other by $\omega_\mathrm{train}$, corresponding to the
repetition rate of the pulses.  As the IP of the He atom is
1.80 Ry, only the high energy wing of these structures shows up in the
PES.  And the agreement between $P_D(\omega)$ and PES is striking for
this part of the spectrum~: Again, both the position and relative
heights of major PES peaks just map the dipole spectrum (mind the
logarithmic ordinate scale and the orders of magnitude suppression of
higher energy PES peaks).  

It is also interesting to note that the dipole power $P_D(\omega)$
(red curve) does not follow exactly the power of the external field
$P_{\rm atto}(\omega)$ (green curve). To understand these discrepancies,
the upper panel shows the pure optical
response $P_D^\mathrm{boost}(\omega)$ obtained from a pure dipole
boost, as that shown in Figure \ref{fig:boostHe}.  This reflects the dipole
eigenmodes of the system.  There are marked differences between $P_{\rm atto}$
and $P_D$ exactly at the places where we observe strong peaks in 
$P_D^\mathrm{boost}$. This indicates that the spectrum of
the dipole response $P_D(\omega)$ combines structures from both
the laser field (i.e. from $P_\mathrm{atto}$) and  the 
strongest eigenmodes of He in $P_D^\mathrm{boost}$. This by
itself is not surprising. What is more interesting is the fact that
the PES precisely matches the combination of both spectra as given in
$P_D$, down to details of peak splittings.  This is especially
clear for very low energy PES peaks in which secondary peaks of
$P_D^\mathrm{boost}(\omega)$ are still visible.  This example of an
attotrain thus confirms our previous findings, namely that
$P_D(\omega\!=\!E_\mathrm{kin}\!-\!\varepsilon_{1s})$ is strongly
related to the PES. The simple estimate remains robust even for an
involved pulse plus admixtures of system frequencies in $P_D$.

\section{Towards a general and robust interpretation of PES}
\label{sec:quad}

\subsection{The surprising case of an attotrain combined to an IR laser}

\begin{figure}[htb!]
 \centerline{\includegraphics[width=0.8\linewidth]{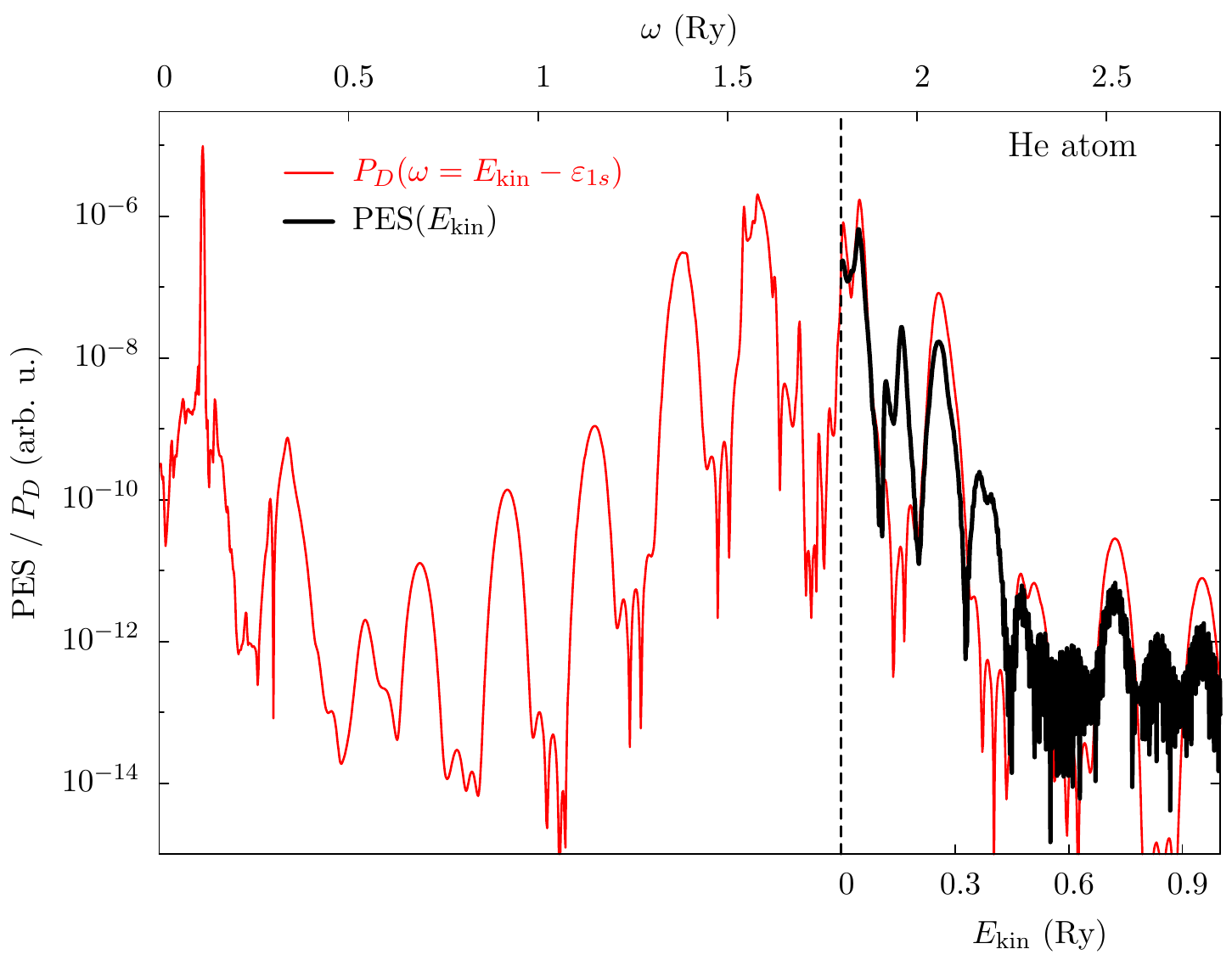}}
 \caption{ PES (black) and dipole power $P_D$ (red) for a He atom
   excited by a combination of an attotrain (\ref{eq:attotrain}) (as used
   in Figure \ref{fig:pureAPT}) with an IR pulse, see text for
   details.  The vertical solid line indicates the ionization
   threshold.}
 \label{fig:IRAPT}
\end{figure}

There remains an interesting case to be explored in connection to
attosecond trains. In most experiments \cite{Joh05,Joh07,Hol11,Klu11}, an IR pulse is used to generate high harmonics which then serve to produce 
the coherent attotrain. Therefore,
the attotrain is shot on top of the IR pulse, the latter one
provoking slow and gentle dipole oscillations of the electron
cloud. We use here an IR pulse with the profile Eq.(\ref{eq:simplepulse})
and the parameters $\omega_\mathrm{las}=\omega_\mathrm{IR}=0.115$~Ry, an overall pulse length $T=T_\mathrm{IR}=105$ fs, and 
a field strength of $E_0=0.0034$ Ry/a$_0$ (that is, an intensity
$I_\mathrm{IR}=10^{11}$ W/cm$^ 2$). The IR pulse alone does not ionize
the ground state of He atom, while the UV attotrain on top of the IR pulse leads to a measurable
ionization. The result of such a setup is shown in Figure
\ref{fig:IRAPT}, again comparing the PES and the dipole power
$P_D$. 
At variance with all previous cases, the two spectra do not
coincide anymore. Only half of the peaks of the
PES can be found in the dipole spectrum $P_D$. 
Another sizeable ingredient is thus clearly missing in this case.

\subsection{Beyond dipole response}

Let us remind that $P_D$
is the Fourier transform of the time evolution of the electronic dipole.
A laser pulse, being $\propto\hat{D}$, therefore excites predominantly the dipole.
However, strong pulses can also induce higher order deformation of the
electron cloud. For example, a dipole shift of the many-electron 
wave function is realized by the 
operation
\begin{equation}
  |\Phi_d\rangle 
  =
  e^{-\mathrm{i}d\hat{P}}|\Phi_0\rangle 
  \approx
  \left(1-\mathrm{i}d\hat{P}-\frac{1}{2}d^2\hat{P}^2\right)
  |\Phi_0\rangle 
\label{eq:deform}
\end{equation}
where $\hat{P}$ is the operator of total electron momentum and $d$ the
size of the shift. Only the term $\propto\hat{P}^1$ is usually
accounted for in case of very weak fields as they are used typically
for nanosecond pulses. Such extremely weak pulses can excite only
modes with odd parity. Pulses in the femtosecond regime, as used here, deal
with stronger (although still moderate) intensities. And this brings
also the term $\propto\hat{P}^2$ into play which, in turn, triggers
modes with even parity. The leading even parity modes are found
in the quadrupole (deformation) and monopole (stretching/compression)
channels.  The corresponding observables are :
\begin{eqnarray}
Q &=&  \int \textrm d \mathbf r\,{\hat Q}\varrho = \int \textrm d \mathbf 
r \, (3z^2 -x^2-y^2)\varrho({\bf r})\\
M &=&  \int \textrm d \mathbf r\,\hat{M}\varrho({\bf r}) =\int \textrm d \mathbf r \, r^2\varrho({\bf r})
\label{eq:mutipoles}
\end{eqnarray}
where $\varrho({\bf r})$ is the local electron density.  Both can be
constructed by angular-momentum reduction of a tensor built from the
square of a dipole $\mathbf{r}\otimes\mathbf{r}$ \cite{Edm57}.
Let us consider for the moment the quadrupole case. Considering the
deformed state defined in Eq.(\ref{eq:deform}), we find
\begin{equation}
  \langle\Phi_d \big|\, \hat{Q} \, \big|\Phi_d\rangle
  =
  \underbrace{\langle\Phi_0 \big| \, \hat{Q} \, \big|\Phi_0\rangle}_{=0}
  -\mathrm{i}d
  \underbrace{\langle\Phi_0 \big| \, [\hat{Q},\hat{P}]\, \big|\Phi_0\rangle}_{=0}
  +
  \frac{d^2}{2}
  \underbrace{\langle\Phi_0 \big| \,[\hat{P},[\hat{Q},\hat{P}]] \, \big|\Phi_0\rangle}_{\neq{0}}
\end{equation}
which shows that the second order term can produce, indeed, a finite
quadrupole momentum.  Whereas the observable $P_D$ can measure only the
odd-parity modes, the PES measuring operator is an outgoing wave,
approximately
$\varphi_\mathrm{k}\propto{e}^{\mathrm{i}\mathbf{k}\cdot\mathbf{r}}$,
which contains odd and even parities. Thus the PES is, in principle,
able to record also even-parity modes. The idea is now that the PES
peaks which are missing in $P_D$ correspond to even-parity excitations
and thus should show up in the power spectrum of the quadrupole $P_Q$
and/or monopole $P_M$.

\subsection{A pure quadrupole excitation as a proof of principle}

Before attempting to analyze the spectrum of Figure \ref{fig:IRAPT}
with even-parity strengths, one should check whether such a picture
makes sense in a ``cleaner\rq\rq{} case avoiding that the quadrupole
contribution has to be figured out from a mix with a dominating
dipole. It would thus be interesting to consider a pure quadrupole
case excluding any dipole. Although it is not clear how to excite a
pure quadrupole in practice, it provides an interesting Gedanken
experiment worth being explored. We thus consider an instantaneous
quadrupole boost to the electron cloud of the He atom, as
\begin{equation}
  \varphi_i({\bf r}) 
  \longrightarrow
  e^{\mathrm{i}\lambda\hat{Q}}\varphi_i(\mathbf{r})
\label{eq:quadboost}
\end{equation}
where $\lambda$ provides the amplitude of the boost. 
The He ground state has even parity and so does the excitation. Thus
the dipole moment
remains zero (within the numerical accuracy) all over the
time evolution and no dipole eigenfrequency is excited in the
process.  Note that the quadrupole boost mostly leads to quadrupole
oscillations, but the pattern are accompanied by small
monopole oscillations which, however, are marginal
for the present case.

\begin{figure}[htb!]
  \centerline{\includegraphics[width=0.8\linewidth]{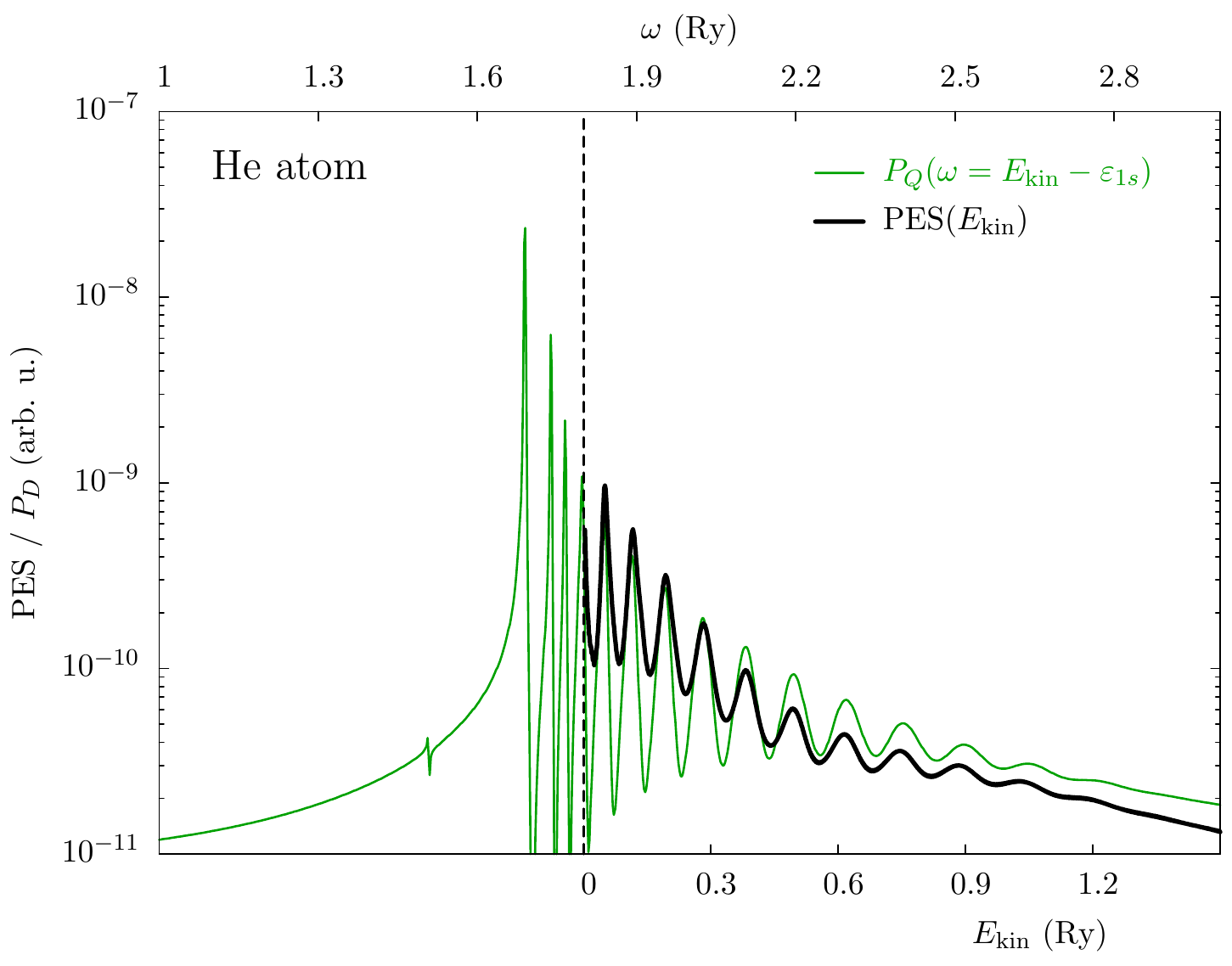}}
 \caption{PES and quadrupole power spectrum $P_Q$ for excitation of the
  He atom by a quadrupole boost, see Eq.~(\ref{eq:quadboost}), 
of amplitude of 0.001/a$_0^2$.}
 \label{fig:quadboost}
\end{figure}

The PES associated to such a quadrupole boost  in a He atom is shown in Figure
\ref{fig:quadboost}. 
We also
plot on the same figure the quadrupole power $P_Q$ which exhibits
the actually excited eigenmodes of the system. The coincidence between
PES and $P_Q$ is again remarkable, both in peak positions and relative
amplitudes. This result tells us that the PES displays the actually
excited eigenfrequencies of the system, whatever way the system is
excited. In the usual case of dipole excitations, the dipole response
is (not surpassingly) exemplified. In a pure quadrupole (no dipole)
excitation, the system mostly responds in the quadrupole channel and
if one performs a monopole excitation, the system dominantly responds
in the monopole channel (not shown here). The actual share of each
mode in the response strongly depends on the excitation but as well on
the system.  As a consequence, one can expect that, for a general
excitation, all channels may be excited (though with different
strengths) and therefore, one should consider all eigenfrequencies on the
same footing. In simple cases, one multipole is dominant so that the
other ones do not show up, but this holds not true in general.

\subsection{Back to the combined  IR + attotrain case}
\begin{figure}[htb!]
  \centerline{\includegraphics[width=0.8\linewidth]{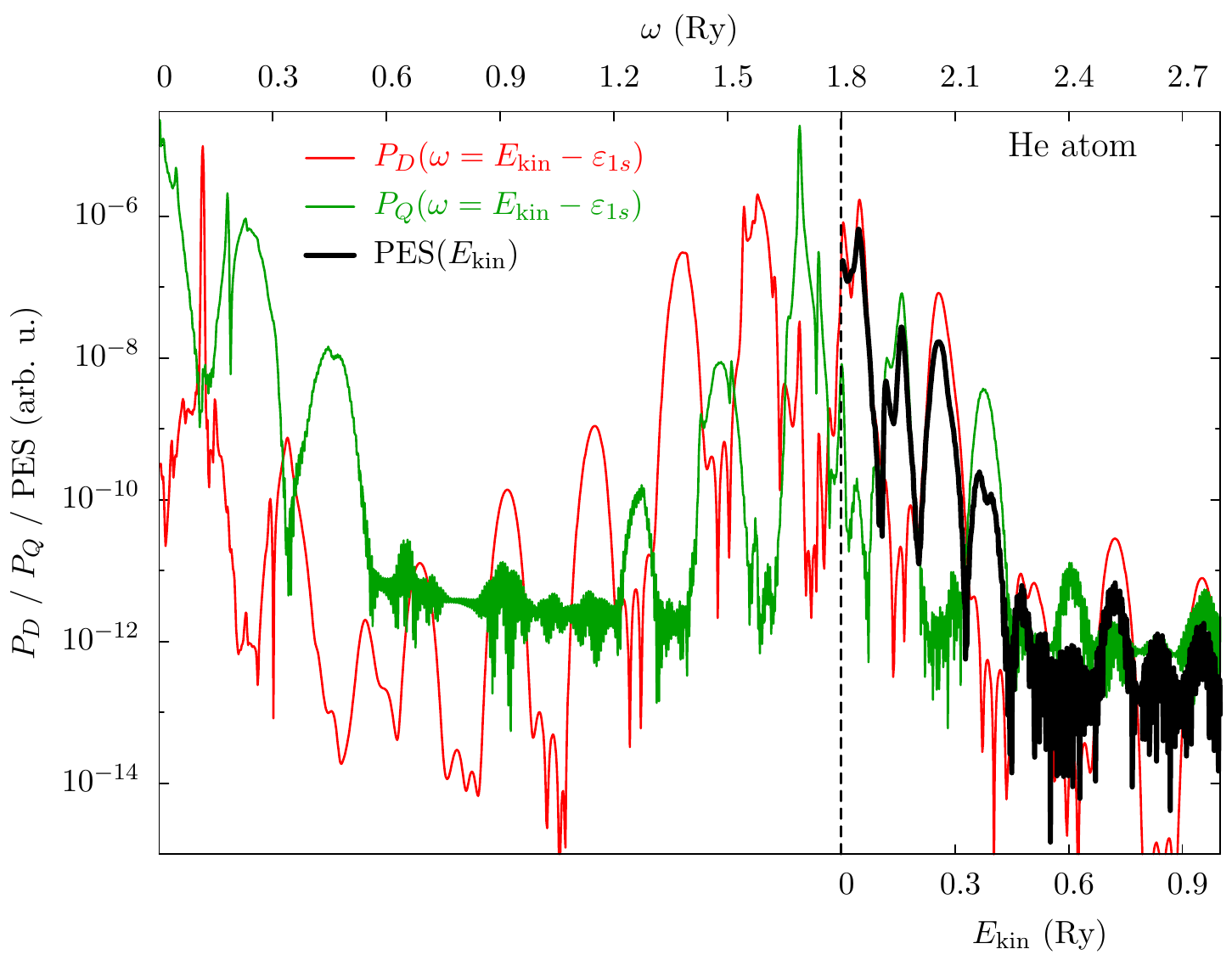}}
 \caption{The same as Fig.~\ref{fig:IRAPT}, but including quadrupole
   power spectra $P_Q$. The $P_D$ and $P_Q$ are scaled with the same
   factor to match the
   height of the PES peaks.}
 \label{fig:IRAPTquad}
\end{figure}

To validate the concept for a more involved case, we come
back to the He atom excited by an IR+APT where the dipole response
alone could not explain the PES, see Figure \ref{fig:IRAPT}.  We now
add the quadrupole spectrum $P_Q$ to the analysis (the monopole
spectrum $P_M$ being two orders of magnitude suppressed compared with $P_Q$, it
is not considered here).  The result is
shown in Figure \ref{fig:IRAPTquad}. 
First note that both $P_D$ and
$P_Q$ take similar values after the {\it same} rescaling to make them
match with the PES scale. This is a clear indication that, in this case,
both dipole and quadrupole channel do respond with comparable
amplitudes. When piling up both $P_D$ and $P_Q$, one finds a perfect
reproduction of the PES, the quadrupole spectrum $P_Q$ precisely
bringing the peaks missed by the dipole spectrum.  Again the two
spectra reproduce both positions and, to a large extent, relative
amplitudes, of the PES peaks. 

\subsection{The Multi-Photon Ionization regime revisited}

\begin{figure}[htb!]
  \centerline{\includegraphics[width=0.8\linewidth]{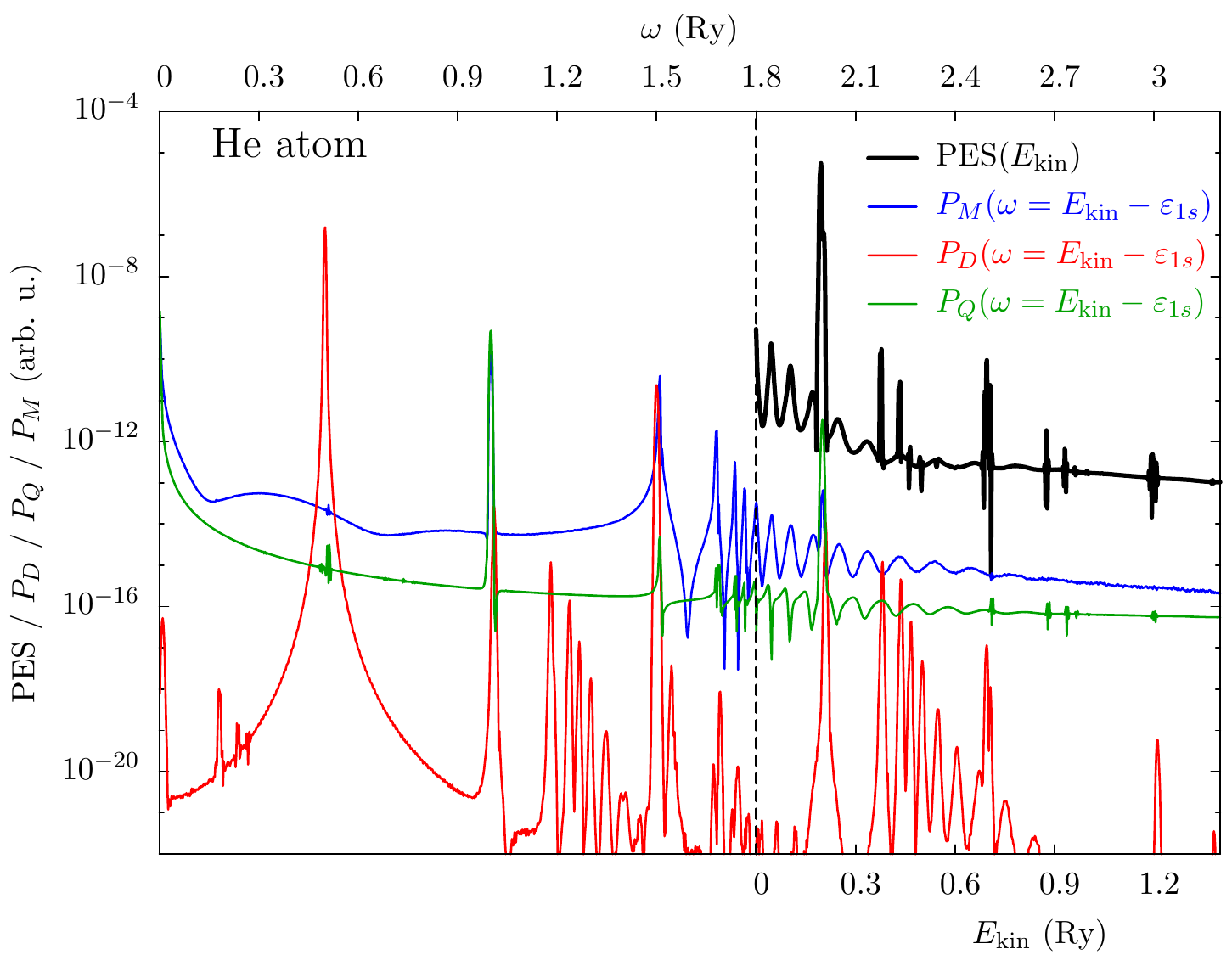}}
 \caption{ PES (black), powers spectra of dipole $P_D$ (red),
   quadrupole $P_Q$ (green), and monopole $P_M$ (blue) for a He atom excited
   by a UV pulse, see Eq.~(\ref{eq:Elas}),
with a frequency of $\omega_{\rm las}=$ 0.5 Ry, intensity 10$^{11}$ W/cm$^{2}$, 
and duration $T=105$~fs.
 }
\label{fig:mpi}
\end{figure}

Another typical scenario in laser driven dynamics is multi-photon
ionization (MPI). We investigate this regime here using a simple laser
pulse Eq. (\ref{eq:simplepulse}) with frequency of $\omega_{\rm las}=$ 0.5 Ry,
intensity $I~=~10^{11}$ W/cm$^{2}$, and duration
$T_\mathrm{pulse}=$ 105 fs.  The frequency, although in the UV regime,
is far below the IP of 1.8~Ry.  It requires at least four photons to lift an
electron into the continuum. This represents a typical MPI case.  The
result is shown in Figure \ref{fig:mpi}.  
It is obvious that the
dipole power $P_D$ alone cannot explain the whole PES.  Remind that a
sufficient pulse intensity is required in MPI to allow for
multi-photon processes which, in turn, makes co-excitation of even
parity modes very likely. The quadrupole power $P_Q$ is sizable, but
$P_D$ and $P_Q$ together cannot explain all pattern of the PES in this
case. Mind that they both have been scaled the same amount for comparison with the PES, but their 
relative values have not been touched (the same holds for the monopole below). 
In fact, $P_Q$ is of minor importance here. There is a sequence
of PES peaks at low $E_\mathrm{kin}$ which stems clearly from the
monopole modes seen in $P_M$. Thus monopole and quadrupole which stay
formally at the same level of importance should always be included
together in the analysis. It is concluded that
the pattern of PES can be reproduced by a proper combination of
monopole, dipole, and quadrupole strength distributions. 

To conclude this discussion of the MPI regime let us mention  the associated 
regime of  above threshold ionization (ATI) 
induced by high infrared intensities with $I>10^{13}$ W/cm$^{2}$~\cite{Fen10,Wop15,Cam00,Hui13}. In this case, the ponderomotive energy $U_p$ induces an energetic shift which can become comparable to the laser frequency~: for 
$\omega_{\text{IR}}=0.11$~Ry, we have $U_p=0.057$~Ry, which is not negligible anymore. For the sake of simplicity we 
avoided, in all the above discussions, to consider cases where the ponderomotive shift was important. 
But in the ATI regime we have to account for this effect. 
This practically does not change our analysis but for the fact that the multipole spectral distributions 
have to be  shifted by $\varepsilon+U_p$ instead
of $\varepsilon$ only. But once this precaution has been taken we find that the present spectral analysis of the ATI PES
perfectly holds, which perfectly supports our above conclusions.

\section{Conclusions and perspectives}
\label{sec:conclusion}

We have discussed in this paper the relations between a photo-electron
spectrum (PES) and spectral strength distributions of the basic
multipole operators (dipole, monopole, quadrupole) for irradiated
atoms or clusters.  In this investigation, we used a He atom as a test
case to have a simple single particle (s.p.) spectrum at the side of
the system.  For the laser pulses, on the other hand, we consider a
variety of scenarios from a simple boost over a one-frequency
pulse to complicated mix of IR and atto-second pulses.

Traditionally, there is a simple rule of thumb which relates the
position of peaks in the PES to the energies of the occupied
s.p. states plus an appropriate multiple of the laser frequency.  The
simple rule (\ref{eq:basic}) relates peak positions with
  s.p. energies, or in other words, the consequent estimate of the PES
  yield, see Eq.~(\ref{eq:y_dirac}), relates peaks with peaks. This
rule cannot be applied to more complex light pulses which embrace a
couple of different frequencies as, e.g., in case of fast collisions
with a charge ion or of the mixed pulses typically used in attosecond
physics.  A natural generalization of it is to model the spectral
distribution of PES by the spectral distribution of dipole power
$P_D(\omega)$ augmented with s.p. energies. This approach is indeed
able to provide a correct picture of PES in some cases. But part of
the PES maxima is missing in others. A closer analysis shows that the
PES is sensitive to excitations with odd partity as well as those with
even parity. However, the dipole spectrum $P_D$ sees only the
odd-parity modes. Therefore one must also extend the estimate by
strengths of even-parity multipoles. The monopole and quadrupole
operator are the most important in that regime. The idea is thus to
include, in addition to $P_D$, the spectral distribution of
monopole power $P_M$ and of quadrupole power $P_Q$ into the
generalized estimate, which amounts to
\begin{equation}
  \mathcal{Y}(E_\mathrm{kin})
  \leftrightarrow
  \sum_i
  \eta_MP_M(E_\mathrm{kin}\!-\!\varepsilon_{i})  
  +
  \eta_DP_D(E_\mathrm{kin}\!-\!\varepsilon_{i})  
  +
  \eta_QP_Q(E_\mathrm{kin}\!-\!\varepsilon_{i})  
\label{eq:general}
\end{equation}
where the sum runs over all occupied s.p. states $i$, and the $\eta$'s factors
take care of the relative strength of each eigenmode.  Experience
gained in the present study suggests that the scaling factors can be
taken as being the same for all modes, i.e.  $\eta_M=\eta_D=\eta_Q$.
This spectrally extended rule (\ref{eq:general}) has been confirmed
for all the types of excitation we have explored, with the contribution of the three terms in
the rule varing from case to case.  This mapping between response(s)
and PES we thus found turns out to be extremely robust.

All in all, the emerging picture remains simple in the sense that the
PES still provides a direct printout of the frequencies contained in
the response of the system.  The simple, traditional, rule
(\ref{eq:basic}) is recovered when exciting by a simple
mono-frequency pulse sufficiently far from any resonance of the irradiated system. This is
the case where we recognize only the laser frequency (and
s.p. energies) in the PES.  The new rule (\ref{eq:general}) covers
much more general situations.  A typical example is attosecond physics
where a complex mix of IR pulse with trains of attosecond pulses 
is often used. There remain widely disputed questions on the
mechanisms underlying the response of the system to such complex
excitations. A typical observable here is the total ionisation which is found to 
oscillate with the delay between IR pulse and attotrain
\cite{Joh05,Nei13,Dah12}. Our preliminary computations show that 
these modulations are (not surprisingly) directly reflected in the
PES. The robust link we have established with multipole response thus
implies a relation between ionisation pattern and multipole
response. This is certainly an interesting connection worth 
being investigated. Work along that line is in progress.\\

\section*{Acknowledgments}

We thank P. Sali\`eres and R. Ta\"ieb for useful discussions.
C.-Z.G. is grateful to the financial support from China Scholarship
Council (CSC) (No. [2013]3009). We thank Institut Universitaire de France, 
European ITN network CORINF and French ANR contract MUSES for 
support during the realization of
this work. It was also granted access to the HPC resources of CalMiP (Calcul en Midi-Pyr\'en\'ees) 
under the allocation P1238, and of RRZE (Regionales Rechenzentrum Erlangen).\\

\bibliographystyle{iopart-num}

\bibliography{library}

\providecommand{\newblock}{}
\begin{thebibliography}{10}
\expandafter\ifx\csname url\endcsname\relax
  \def\url#1{{\tt #1}}\fi
\expandafter\ifx\csname urlprefix\endcsname\relax\def\urlprefix{URL }\fi
\providecommand{\eprint}[2][]{\url{#2}}

\bibitem{Tur62a}
Turner D~W and Jobory M~I~A 1962 {\em J. Chem. Phys.\/} {\bf 37} 3007

\bibitem{Tur70aB}
Turner D 1970 {\em Molecular Photoelectron Spectroscopy\/} (New York: Wiley)

\bibitem{Mai91b}
Mainfray G and Manus G 1991 {\em Reports on progress in physics\/} {\bf 54}
  1333

\bibitem{Tje90a}
Tjeng L~H, Vos A~R and Sawatzky G~A 1990 {\em Surf. Sci.\/} {\bf 235} 269

\bibitem{Hof01}
Hoffmann M~A, Wrigge G, v~Issendorff B, Muller J, Gantefor G and Haberland H
  2001 {\em Eur. Phys. J. D\/} {\bf 16} 9

\bibitem{Liu98a}
Liu H and Hamers R~J 1998 {\em Surface Science\/} {\bf 416} 354

\bibitem{McH89}
McHugh K~M, Eaton J~G, Lee G~H, Sarkas H~W, Kidder L~H, Snodgrass J~T, Manaa
  M~R and Bowen K~H 1989 {\em J. Chem. Phys.\/} {\bf 91} 3792

\bibitem{Rab77aB}
Rabalais J 1977 {\em Principles of Ultraviolet Photoelectron Spectroscopy\/}
  (New York: Wiley)

\bibitem{Fai87}
Faisal F~H~M 1987 {\em {Theory of Multiphoton Processes}\/} (New York: Plenum
  Press)

\bibitem{Fen10}
Fennel T, Meiwes-Broer K~H, Tiggesb\"aumker J, Dinh P~M, Reinhard P~G and
  Suraud E 2010 {\em Rev. Mod. Phys.\/} {\bf 82} 1793

\bibitem{Ago79}
Agostini P, Fabre F, Mainfray G, Petite G and Rahman N~K 1979 {\em Phys. Rev.
  Lett.\/} {\bf 42}(17) 1127--1130

\bibitem{Ebe91}
Eberly J~H, Javanainen J and Rza̧{\.z}ewski K 1991 {\em Physics reports\/}
  {\bf 204} 331--383

\bibitem{DeW98}
DeWitt M~J and Levis R~J 1998 {\em Phys. Rev. Lett.\/} {\bf 81}(23) 5101--5104

\bibitem{Cam01}
Campbell E, Hoffmann K, Rottke H and Hertel I 2001 {\em Journal of Chemical
  Physics\/} {\bf 114} 1716--1719

\bibitem{Hat07}
Hatamoto T, Okunishi M, Lischke T, Pr{\"u}mper G, Shimada K, Mathur D and Ueda
  K 2007 {\em Chemical physics letters\/} {\bf 439} 296--300

\bibitem{Kel65}
Keldysh L~V 1965 {\em Sov. Phys. JETP\/} {\bf 20} 1307

\bibitem{Kel03}
Keller U 2003 {\em Nature\/} {\bf 424} 831

\bibitem{Rul05}
Rulli\`ere C (ed) 2005 {\em {Femtosecond Laser Pulses: Principles and
  Experiments, 2nd ed., Advanced Texts in Physics}\/} (New-York: Springer)

\bibitem{Pas08}
Paschotta R 2008 {\em {Encyclopedia of Laser Physics and Technology, volumes 1
  and 2}\/} (Berlin: Wiley-VCH)

\bibitem{Zew94}
Zewail A~H 1994 {\em {Femtochemistry, Vol. I \& II}\/} (Singapore: World
  Scientific)

\bibitem{Sei00}
Seifert G, Kaempfe M, Berg K~J and Graener H 2000 {\em Appl. Phys. B\/} {\bf
  71} 795

\bibitem{And02}
Andrae K, Reinhard P~G and Suraud E 2002 {\em J. Phys. B\/} {\bf 35} 1

\bibitem{Pau01}
Paul P~~M, Toma E, Breger P, Mullot G, Aug{\'e} F, Balcou P, Muller H and
  Agostini P 2001 {\em Science\/} {\bf 292} 1689--1692

\bibitem{Kra09}
Krausz F and Ivanov M 2009 {\em Rev. Mod. Phys.\/} {\bf 81}(1) 163--234

\bibitem{Cor07}
Corkum P and Krausz F 2007 {\em Nature Physics\/} {\bf 3} 381--387

\bibitem{Joh05}
Johnsson P, L\'opez-Martens R, Kazamias S, Mauritsson J, Valentin C, Remetter
  T, Varj\'u K, Gaarde M~B, Mairesse Y, Wabnitz H, Sali\`eres P, Balcou P,
  Schafer K~J and L'Huillier A 2005 {\em Phys. Rev. Lett.\/} {\bf 95}(1) 013001

\bibitem{Nei13}
Neidel C, Klei J, Yang C~H, Rouz\'ee A, Vrakking M~J~J, Kl\"under K, Miranda M,
  Arnold C~L, Fordell T, L'Huillier A, Gisselbrecht M, Johnsson P, Dinh M~P,
  Suraud E, Reinhard P~G, Despr\'e V, Marques M~A~L and L\'epine F 2013 {\em
  Phys. Rev. Lett.\/} {\bf 111}(3) 033001

\bibitem{Kel10}
Kelkar A, Kadhane U, Misra D, Gulyas L and Tribedi L 2010 {\em Phys. Rev. A\/}
  {\bf 82}(4) 043201

\bibitem{Nan13}
Nandi S, Biswas S, Khan A, Monti J, Tachino C, Rivarola R, Misra D and Tribedi
  L 2013 {\em Phys. Rev. A\/} {\bf 87}(5) 052710

\bibitem{Din15}
Dinh P~M, Reinhard P~G, Suraud E and Wopperer P 2015 {\em Eur. Phys. J. D\/}
  {\bf 69} 48

\bibitem{Hee93}
de~Heer W~A 1993 {\em Rev. Mod. Phys.\/} {\bf 65} 611

\bibitem{Bra93}
Brack M 1993 {\em Rev. Mod. Phys.\/} {\bf 65} 677

\bibitem{Her06aR}
{Hertel} I~V and {Radloff} W 2006 {\em Rep. Prog. Phys.\/} {\bf 69} 1897

\bibitem{Poh01}
Pohl A, Reinhard P~G and Suraud E 2001 {\em J. Phys. B\/} {\bf 34} 4969

\bibitem{Koh65}
Kohn W and Sham L~J 1965 {\em Phys. Rev.\/} {\bf 140} 1133

\bibitem{Gro90}
Gross E~K~U and Kohn W 1990 {\em Adv. Quant. Chem.\/} {\bf 21} 255

\bibitem{Kue98}
K\"ummel S, Brack M and Reinhard P~G 1998 {\em Phys. Rev. B\/} {\bf 58} 1774

\bibitem{Goe96}
Goedecker S, Teter M and Hutter J 1996 {\em Phys. Rev. B\/} {\bf 54} 1703

\bibitem{Per92}
Perdew J~P and Wang Y 1992 {\em Phys. Rev. B\/} {\bf 45} 13244

\bibitem{Leg02}
Legrand C, Suraud E and Reinhard P~G 2002 {\em J. Phys. B\/} {\bf 35} 1115

\bibitem{Klu13a}
Kluepfel P, Dinh P~M, Reinhard P~G and Suraud E 2013 {\em Phys. Rev. A\/} {\bf
  88} 052501

\bibitem{Wop15}
Wopperer P, Dinh P~M, Reinhard P~G and Suraud E 2015 {\em Phys. Rep.\/} {\bf
  562} 1

\bibitem{Joh07}
Johnsson P, Mauritsson J, Remetter T, L'Huillier A and Schafer K~J 2007 {\em
  Phys. Rev. Lett.\/} {\bf 99}(23) 233001

\bibitem{Hol11}
Holler M, Schapper F, Gallmann L and Keller U 2011 {\em Phys. Rev. Lett.\/}
  {\bf 106} 123601

\bibitem{Klu11}
Kl\"under K, Dahlstr\"om J~M, Gisselbrecht M, Fordell T, Swoboda M, Gu\'enot D,
  Johnsson P, Caillat J, Mauritsson J, Maquet A, Ta\"ieb R and L'Huillier A
  2011 {\em Phys. Rev. Lett.\/} {\bf 106} 143002

\bibitem{Cal00}
Calvayrac F, Reinhard P~G, Suraud E and Ullrich C~A 2000 {\em Phys. Rep.\/}
  {\bf 337} 493

\bibitem{Rei03a}
Reinhard P~G and Suraud E 2003 {\em Introduction to Cluster Dynamics\/} (New
  York: Wiley)

\bibitem{Rei06c}
Reinhard P~G, Stevenson P~D, Almehed D, Maruhn J~A and Strayer M~R 2006 {\em
  Phys. Rev. E\/} {\bf 73} 036709

\bibitem{Cal95a}
Calvayrac F, Reinhard P~G and Suraud E 1995 {\em Phys. Rev. B\/} {\bf 52}
  R17056

\bibitem{Cal97b}
Calvayrac F, Reinhard P~G and Suraud E 1997 {\em Ann. Phys. (NY)\/} {\bf 255}
  125

\bibitem{Poh04a}
Pohl A, Reinhard P~G and Suraud E 2004 {\em J. Phys. B\/} {\bf 37} 3301

\bibitem{Bae10}
B\"ar M, Dinh P~M, moskaleva L~V, Reinhard P~G, R\"osch N and Suraud E 2010
  {\em Phys. Stat. Sol. B\/} {\bf 247} 989

\bibitem{DeG12}
De~Giovannini U, Varsano D, Marques M~A~L, Appel H, Gross E~K~U and Rubio A
  2012 {\em Phys. Rev. A\/} {\bf 85}(6) 062515

\bibitem{Din13}
Dinh P~M, Romaniello P, Reinhard P~G and Suraud E 2013 {\em Phys. Rev. A\/}
  {\bf 87} 032514

\bibitem{Wop12c}
Wopperer P, Reinhard P~G and Suraud E 2013 {\em Ann. Phys. (Berlin)\/} {\bf
  525} 309--321

\bibitem{Edm57}
Edmonds A~R 1957 {\em {Angular Momentum in Quantum Mechanics}\/} (Princeton:
  Princeton University Press)

\bibitem{Cam00}
Campbell E~E~B, Hansen K, Hoffmann K, Korn G, Tchaplyguine M, Wittmann M and
  Hertel I~V 2000 {\em Phys. Rev. Lett.\/} {\bf 84} 2128

\bibitem{Hui13}
Huismans Y, Cormier E, Cauchy C, Hervieux P~A, Gademann G, Gijsbertsen A,
  Ghafur O, Johnsson P, Logman P, Barillot T, Bordas C, L\'epine F and Vrakking
  M~J~J 2013 {\em Phys. Rev. A\/} {\bf 88}(1) 013201

\bibitem{Dah12}
Dahlstr{\"o}m J, L’Huillier A and Maquet A 2012 {\em Journal of Physics B:
  Atomic, Molecular and Optical Physics\/} {\bf 45} 183001

\end{thebibliography}

\end{document}